\documentclass[amsmath,amsfonts,amssymb,twocolumn,nofootinbib]{revtex4-1}
\usepackage{graphicx}
\usepackage{verbatim}
\usepackage{hyperref}
\usepackage{epstopdf}
\usepackage{subfigure}

\begin{document}

\title{Generalizing the No-U-Turn Sampler to Riemannian Manifolds}
\author{Michael Betancourt}
\affiliation{Applied Statistics Center, Columbia University, New York, NY 10027, USA}
\email{betanalpha@gmail.com}


\begin{abstract}

Hamiltonian Monte Carlo provides efficient Markov transitions at the expense of introducing two free parameters: a step size and total integration time.  Because the step size controls discretization error it can be readily tuned to achieve certain accuracy criteria, but the total integration time is left unconstrained.  Recently Hoffman and Gelman proposed a criterion for tuning the integration time in certain systems with their No U-Turn Sampler, or NUTS.  In this paper I investigate the dynamical basis for the success of NUTS and generalize it to Riemannian Manifold Hamiltonian Monte Carlo.

\end{abstract}

\maketitle

Taking advantage of a natural mapping between forms on a manifold and measures on a probability space, Hamiltonian Monte Carlo (HMC) generates Metropolis proposals by simulating Hamiltonian flow.  The autocorrelation of the resulting Markov chain depends on how long the flow is integrated, but there are no conditions on the optimal integration time that minimizes autocorrelation for a general target distribution.  When the target distribution is unimodal, however, there is a natural criterion: turning points of the resultant trajectories.  The No U-Turn Sampler identifies these turning points for Euclidean manifolds, but the criterion begins to fail when applied to more complex distributions and Riemannian manifolds.  Appealing to the geometry of HMC, however, admits a straightforward generalization of the No U-Turn Sampler that is not only amenable to Riemannain manifolds but also isolates the turning points in more complicated, non-convex target distributions.

\section*{Hamiltonian Monte Carlo}

Given a target density $\pi \! \left( \mathbf{q} \right)$, Hamiltonian Monte Carlo~\cite{Duane1987, Neal2011, Betan2011} yields Metropolis proposals for the extended density,
\begin{equation*}
\pi \! \left( \mathbf{p}, \mathbf{q} \right) = \exp \left[ - H \! \left( \mathbf{p}, \mathbf{q} \right) \right],
\end{equation*}
where the Hamiltonian, $H \! \left( \mathbf{p}, \mathbf{q} \right)$, is defined as
\begin{equation*}
H \! \left( \mathbf{p}, \mathbf{q} \right) = T \! \left( \mathbf{p}, \mathbf{q} \right) + V \! \left( \mathbf{q} \right),
\end{equation*}
with the potential energy, $V \! \left( \mathbf{q} \right)$ defined by the target density,
\begin{equation*}
V \! \left( \mathbf{q} \right) = - \log \pi \! \left( \mathbf{q} \right).
\end{equation*}
Up to a few weak constraints, the kinetic energy, $T \! \left( \mathbf{p}, \mathbf{q} \right)$ can to tuned to improve the proposals.

In particular, HMC defines the potential as a function on some manifold $\mathcal{M}$ with coordinate functions $\mathbf{q}$, and the Hamiltonian as scalar function on the cotangent bundle, $T^{*} \mathcal{M}$, with coordinate functions $\left\{ \mathbf{p}, \mathbf{q} \right\}$.  At a point $S \in T^{*} \mathcal{M}$ the Hamiltonian is then more formally written as
\begin{equation*}
H \! \left( S \right) = T \! \left( S \right) + V \! \left( R \right),
\end{equation*}
where $R = \pi \! \left( S \right) \in \mathcal{M}$\footnote{Here $\pi$ is the natural  projection operator $\pi: T^{*} \mathcal{M} \rightarrow \mathcal{M}$ on the cotangent bundle, not the notation for probability density used above.  A point $R \in \mathcal{M}$ is identified by the same position coordinate functions $\mathbf{q}$ in $T^{*} \mathcal{M}$, so that $\mathbf{q} \! \left( R \right) = \mathbf{q} \! \left( S \right)$.  The value of the momentum coordinate functions identify a unique covector on the base manifold, $\mathbf{p} \! \left( R \right) \in T^{*}_{R} \mathcal{M}$.}.

Given the natural symplectic geometry of $T^{*} \mathcal{M}$, the Hamiltonian defines a vector field, and integrating along the vector field for some time $t$ generates a flow that maps any point, $S_{0} \in T^{*} \mathcal{M}$, to some final point, $S_{t} \in T^{*} \mathcal{M}$, that serves as the proposal.  

Integrating for only a short time yields highly correlated points and random walk behavior, but trajectories that are integrated for too long may double back on themselves and waste computational resources.  In order to optimize the proposal, we need to determine exactly how long to evolve the trajectories in order to maximize the distance between proposals and minimize the autocorrelation of the chain.

\section*{Harmonic Motion}

In many applications, a Markov chain is meant to sample from a single mode of the target distribution, which manifests as a well in the potential energy.  For small enough values of Hamiltonian, the trajectories become bound in the neighborhood of the mode and execute harmonic motion.

For example, consider \textit{Euclidean Manifold Hamiltonian Monte Carlo} (EMHMC),
\begin{equation*}
H \! \left( S \right) = \frac{1}{2} \sum_{jk} p_{j} \! \left( S \right) p_{k} \! \left( S \right) \left( M^{-1} \right) ^{jk} + V \! \left(R \right),
\end{equation*}
with a quadratic potential,
\begin{equation*}
V \! \left( R \right) = \frac{1}{2} \sum_{jk} q^{j} \! \left( R \right) q^{k} \! \left( R  \right) W_{jk}.
\end{equation*}
The trajectories from this Hamiltonian execute simple harmonic motion~\cite{Jose1998} along the the directions $\mathbf{N}_{j} = \sqrt{\mathbf{M}} \cdot \hat{\mathbf{n}}_{j}$, where the $\hat{\mathbf{n}}_{j}$ are the eigenvectors of $\sqrt{\mathbf{M^{-1}}} \cdot \mathbf{W} \cdot \sqrt{\mathbf{M^{-1}}}$. In terms of the coordinate functions\footnote{The following uses complex exponential notation with an implicit projection to the real axis assumed.  For example, $\exp \left[ i \omega t \right]$ formally represents $\mathcal{R} \left( \exp \left[ i \omega t \right] \right) = \mathcal{R} \left( \cos \left( \omega t \right) + i \sin \left( \omega t \right) \right) = \cos \left( \omega t \right)$.},
\begin{align}
\mathbf{q} \! \left(S_t\right) &= A \sum_{j} \exp \left[ i \left( \omega_{j} t + \phi_{j} + \pi / 2 \right) \right] \mathbf{N}^{j} \\
\mathbf{p} \! \left(S_t\right) &= A \sum_{j} \exp \left[ i \left( \omega_{j} t + \phi_{j} \right) \right] \mathbf{N}_{j}. \label{sho}
\end{align}
with the frequencies given by the respective eigenvalues, $\omega_{j} = \sqrt{\lambda_{j}}$, and the amplitude, $A$, and phases, $\phi_{j}$, determined by the initial conditions.  

The motion along each direction $\mathbf{N}_{j}$ turns back on itself after reaching a \textit{turning point}, where the projection of the momentum vanishes in the process of changing sign.  Because the extent of the trajectory is limited by the oscillation with the longest period, $T_{j} = 2\pi / \omega_{j}$, maximizing distance is equivalent to integrating until reaching the \textit{turning point of the longest oscillation}, or TPOLO for succintness.  For a more general Hamiltonian the motion will be more complex but still harmonic, and the TPOLO still determines the optimal integration time.   

\begin{figure*}[t]
\centering
\subfigure[]{\includegraphics[width=3.5in]{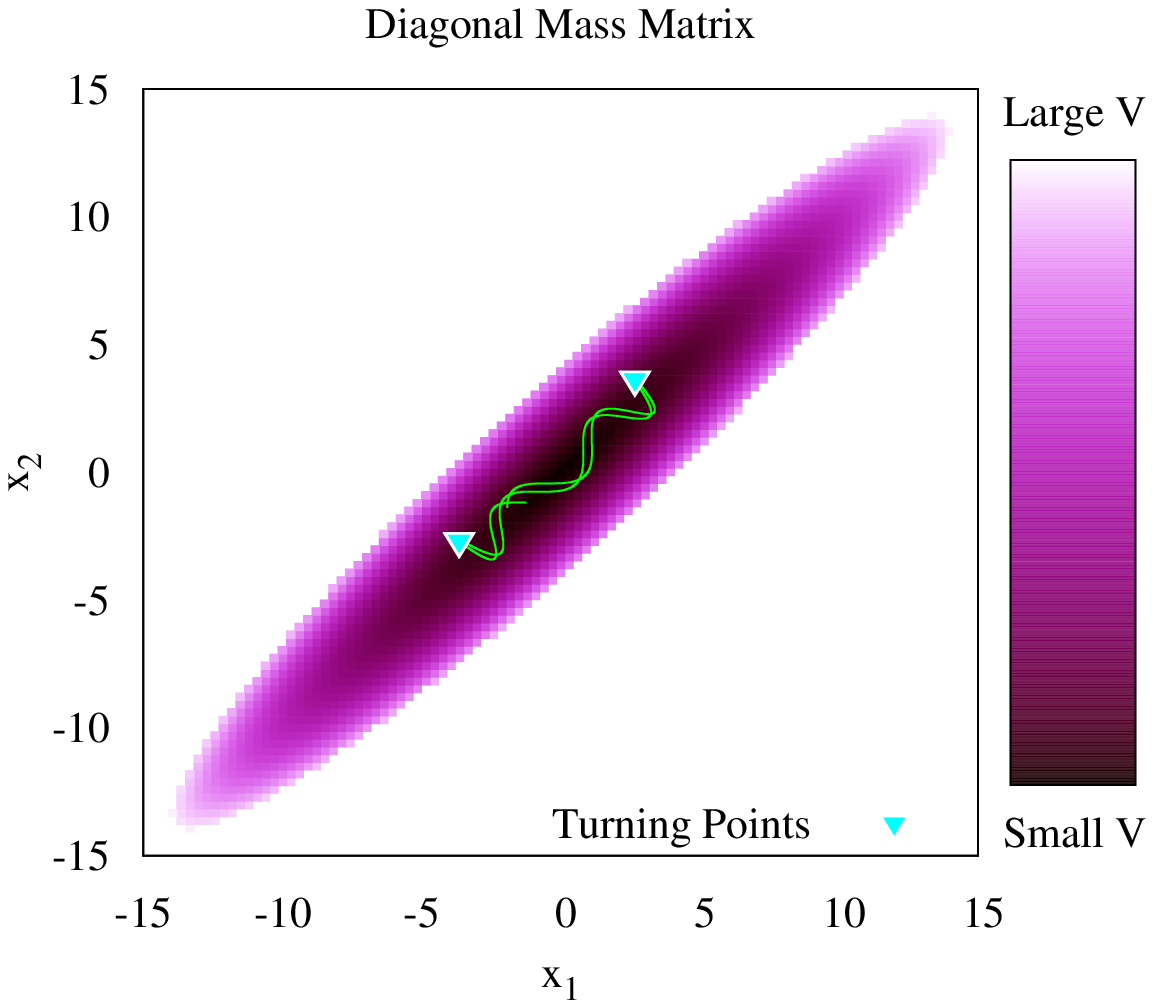}}
\subfigure[]{\includegraphics[width=3.5in]{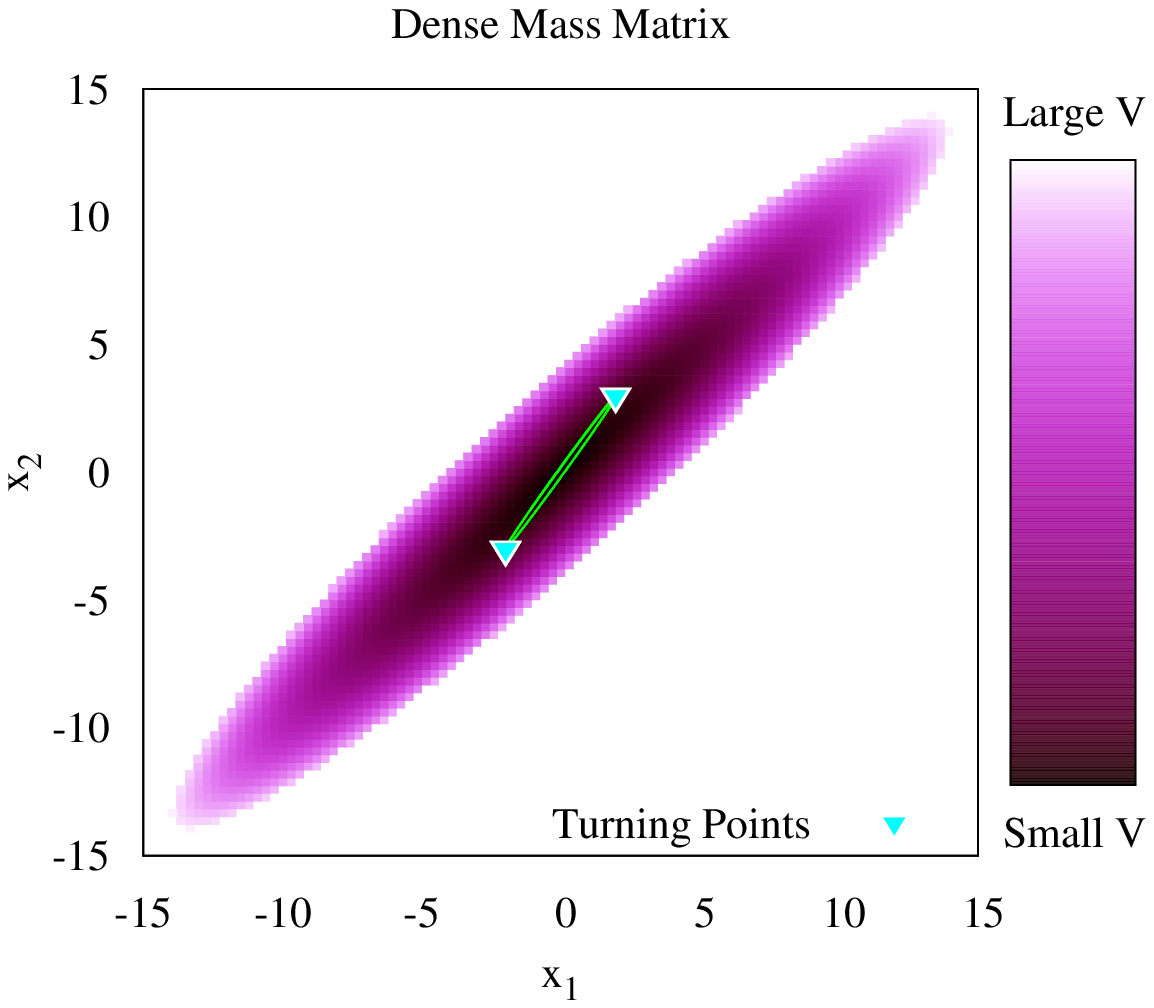}}
\caption{Given the quadratic potential \eqref{gaussV}, the Hamiltonian trajectories execute simple harmonic motion with (a) eigenfrequencies $\omega = \left( 1 \pm \rho \right)^{-\frac{1}{2}}$ for $\mathbf{M} = \mathbb{I}$ and (b) degenerate eigenfrequencies $\omega = 1$ for $\mathbf{M} = \mathbf{\Sigma}^{-1}$, with phases and amplitude determined by the initial position and momentum.  The trajectories turn back on themselves at the turning points of the longest oscillation, which align with the semi-major axis of the potential.
\label{fig:gaussSHO}}
\end{figure*}

Sampling from a Gaussian distribution $\mathcal{N} \! \left( \mathbf{0}, \mathbf{\Sigma} \right)$ with

\begin{equation*}
\mathbf{\Sigma} = \left( \begin{array}{cc} 1 & \rho \\ \rho & 1 \\ \end{array} \right),
\end{equation*}
and $\rho = 0.95$, for example, yields the potential
\begin{equation} \label{gaussV}
V \! \left( R \right) = \frac{1}{2} \sum_{jk} q^{j} \! \left( R \right) q^{k} \! \left( R \right) \left( \Sigma^{-1} \right)_{ij}.
\end{equation}
The resulting Hamiltonian dynamics execute simple harmonic motion, with the bulk of the motion turning back on itself at the TPOLOs (Figure \ref{fig:gaussSHO}).

Identifying the slowest turning point during a trajectory, however, is not an easy task.

\section*{The No U-Turn Sampler}

In HMC, naively terminating a trajectory based on some dynamic criterion sacrifices detailed balance and threatens convergence to the target distribution.  The No U-Turn Sampler, or NUTS, admits a dynamic termination criterion that preserves detailed balance by using Hamiltonian dynamics to build a binary tree from which the transition is sampled~\cite{Hoffman2011}.  

The termination criterion itself attempts to identify the longest turning point by taking advantage of the geometry of EMHMC.  Here the base manifold, $\mathcal{M}$, is endowed with a Euclidean geometry where the distance between two points, $\mathbf{r} \! \left( R_{1}, R_{2} \right) = \mathbf{q} \! \left( R_{1} \right) - \mathbf{q} \! \left( R_{2} \right)$, is a well-defined vector.  Because the momentum is tangential to the trajectory when $\mathbf{M} = \mathbb{I}$, the contraction of the distance onto the current momentum vanishes\footnote{Here we're taking advantage of the fact that a point $S \in T^{*} \mathcal{M}$ identifies both a point, $R = \pi \! \left( S \right) \in \mathcal{M}$ and a covector, $\mathbf{p} \! \left( R \right) \in T^{*}_{R} \mathcal{M}$, to construct the NUTS criterion on the base manifold, $\mathcal{M}$.}, 
\begin{equation*}
\sum_{j} p_{j} \! \left( R_{t} \right) \, r^{j} \! \left( R_{t}, R_{0} \right) = 0,
\end{equation*}
exactly when the trajectory should have started to turn back on itself, and consequently provides a valid termination criterion.  

In practice, the success of the NUTS criterion is not limited to the case of a unit mass matrix.  Consider the expansion of the distance as an integral over the trajectory,
\begin{align*}
\mathbf{r} \! \left( R_{t}, R_{0} \right) &= \mathbf{q} \! \left( R_{t} \right) - \mathbf{q}_{i} \! \left( R_{0} \right) \\
&= \int_{0}^{t} \mathrm{d} \tau \, \mathbf{M}^{-1} \mathbf{p} \! \left(R_\tau \right) \\
&= \mathbf{M}^{-1} \int_{0}^{t} \mathrm{d} \tau \,  \mathbf{p} \! \left(R_\tau \right) \\
&= t \, \mathbf{M}^{-1} \mbox{\boldmath{$\rho$}} \! \left( R_{t} \right),
\end{align*}
where
\begin{align*}
\mbox{\boldmath{$\rho$}} \! \left(R_t\right) \equiv \frac{1}{t} \int_{0}^{t} \mathrm{d} \tau \, \mathbf{p} \! \left( R_\tau \right).
\end{align*}

In terms of $\mbox{\boldmath{$\rho$}} \! \left(R_t\right)$, the NUTS criterion becomes
\begin{align*}
\sum_{j} p_{j}  \! \left(R_t\right) \, r^{j}  \! \left( R_{t}, R_{0} \right) &= 0 \\
t \sum_{jk} p_{j}  \! \left(R_t\right) \rho_{k}  \! \left(R_t\right) \left( M^{-1} \right)^{jk}  &= 0\\
\sum_{jk} p_{j}  \! \left(R_t\right) \rho_{k} \! \left(R_t\right)  \left( M^{-1} \right)^{jk}  &= 0\\
\left< \mathbf{p}  \! \left(R_t\right), \mbox{\boldmath{$\rho$}}  \! \left(R_t\right) \right>_{\mathbf{M}^{-1}}  &= 0,
\end{align*}
where the inner product on $\mathcal{M}$ is defined by
\begin{equation*}
\left< \mathbf{a}, \mathbf{b} \right>_{ \mathbf{\Lambda} } \equiv \sum_{jk} a_{j} b_{k} \Lambda^{jk}.
\end{equation*}

Now, for simple harmonic motion \eqref{sho} ,
\begin{align*}
\mbox{\boldmath{$\rho$}}  \! \left(R_t\right) &= \frac{1}{t} \int_{0}^{t} \mathrm{d} \tau \, A \sum_{j} \exp \left[ i \left( \omega_{j} t + \phi_{j} \right) \right] \mathbf{N}_{j} \\
&= A \sum_{j} \frac{1}{t} \int_{0}^{t} \mathrm{d} \tau \, \exp \left[ i \left( \omega_{j} t + \phi_{t} \right) \right] \mathbf{N}_{j} \\
&= A \sum_{j} \frac{ \exp \left[ i \left( \omega_{j} t + \phi_{j} \right) \right] - \exp \left[ i \phi_{j} \right] }{i \omega_{j} t} \mathbf{N}_{j}.
\end{align*}
so that\footnote{Note that the $\mathbf{N}_{j}$ are orthogonal with respect to the metric $\mathbf{M}^{-1}$.}
\begin{align*}
0 &= \left< \mathbf{p}  \! \left(R_t\right), \mbox{\boldmath{$\rho$}}  \! \left(R_t\right) \right>_{\mathbf{M}^{-1}} \\
&= A^{2} \sum_{jk} \frac{ 1 }{i \omega_{j} t} \left( \exp \left[ i \left( \omega_{j} t + \phi_{j} \right) \right] - \exp \left[ i \phi_{j} \right] \right) \\
& \qquad \qquad \; \times \exp \left[ i \left( \omega_{k} t + \phi_{k} \right) \right] \left< \mathbf{N}_{j}, \mathbf{N}_{k} \right>_{\mathbf{M}^{-1}} \\
&= A^{2} \sum_{j} \frac{ \left( \exp \left[ i \left( \omega_{j} t + \phi_{j} \right) \right] - \exp \left[ i \phi_{j} \right] \right) \exp \left[ i \left( \omega_{j} t + \phi_{j} \right) \right]  }{i \omega_{j} t} \\
&= A^{2} \sum_{j} \frac{ \exp \left[ i \phi_{j} \right] }{i \omega_{j} t} \left( \exp \left[ i \omega_{j} t \right] - 1 \right) \exp \left[ i \left( \omega_{j} t + \phi_{j} \right) \right].
\end{align*}

As the trajectory evolves higher frequency components begin to decay as $\omega_{j} t \gg 1$, isolating the slowest oscillation or at least a set of slow oscillations with degenerate frequencies\footnote{In practice, the transient behavior can lead to premature satisfaction of the criterion;  care must be taken when trajectories are terminated quickly.}.  In either case the criterion reduces to
\begin{align*}
0 &= \left< \mathbf{p}  \! \left(R_t\right), \mbox{\boldmath{$\rho$}}  \! \left(R_t\right) \right>_{\mathbf{M}^{-1}} \\
&=  \frac{A^{2}}{i \omega t} \left( \exp \left[ i \omega t \right] - 1 \right) \exp \left[ i\omega t \right] \sum_{j} \exp \left[ i 2 \phi_{j} \right].
\end{align*}

The first term in parentheses vanishes after every complete oscillation; the more interesting zero occurs when
\begin{equation*}
\exp \left[ i\omega t \right] \sum_{j} \exp \left[ i 2 \phi_{j} \right] = 0.
\end{equation*}
If the phases are coherent, $\phi_{j} = \phi \, \forall j$, then the zero occurs at the nearest turning point,
\begin{equation*}
t = \frac{T}{2} \left[ n + \left(1 - \frac{2 \phi }{\pi} \right) \right], \, n \in \mathbb{Z},
\end{equation*}
but if the phases are incoherent then the total momentum is constant and there are no unique turning points along the trajectory.  In this case the sum tends towards unity leaving the zero at
\begin{equation*}
t = \frac{T}{2},
\end{equation*}
which maximizes the distance by integrating to the opposite point of the orbit.

\begin{figure*}[]
\centering
\includegraphics[width=7in]{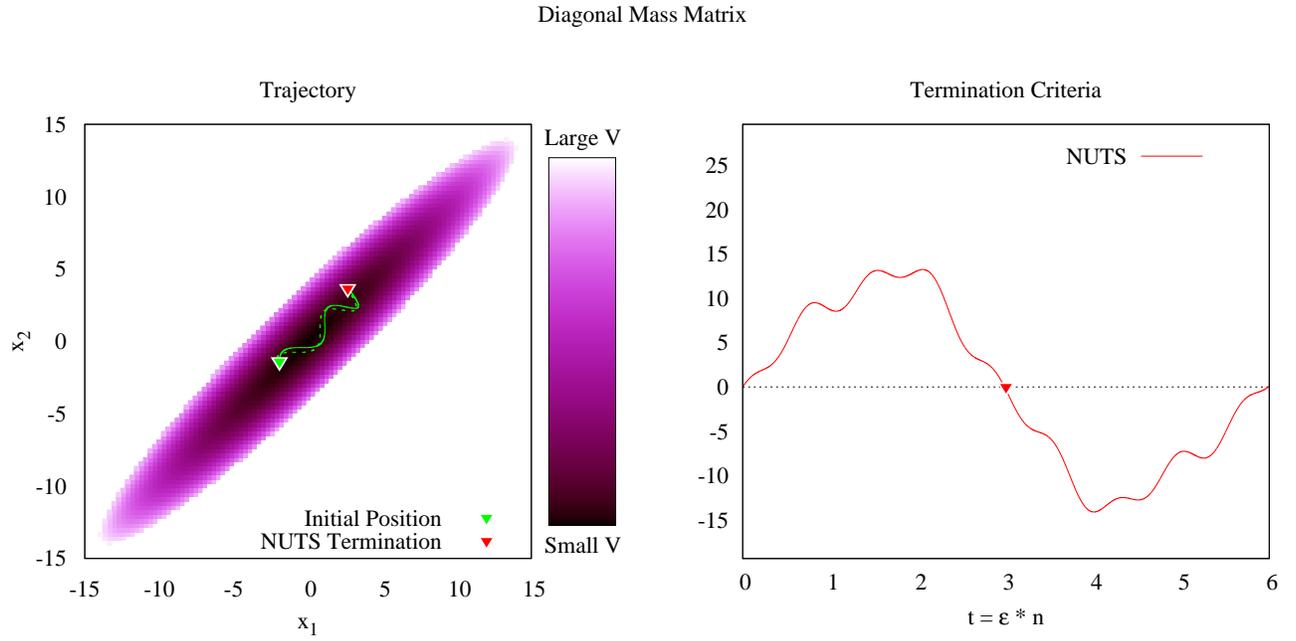}
\caption{When applied to the quadratic potential \eqref{gaussV} with $\mathbf{M} = \mathbb{I}$ (Figure \ref{fig:gaussSHO}a), the NUTS criterion terminates the trajectory at the first TPOLO, as desired.
\label{fig:gaussNUTS1}}
\vspace{5mm}
\end{figure*}

\begin{figure*}[]
\centering
\includegraphics[width=7in]{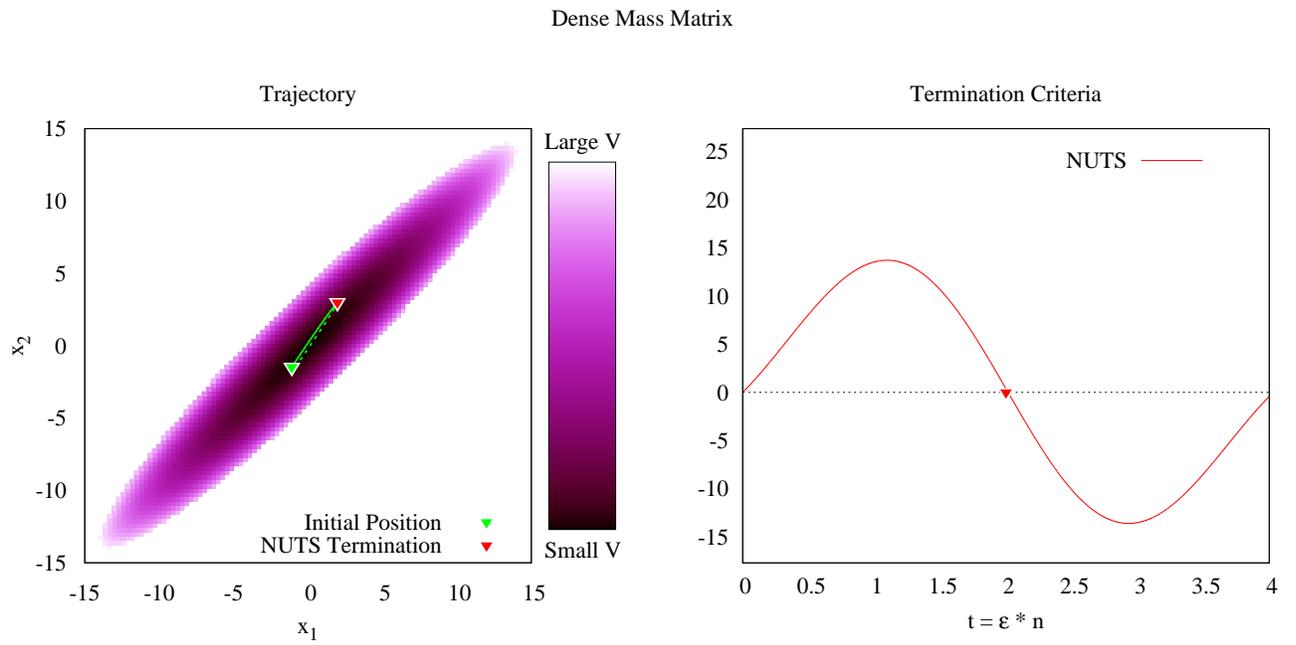}
\caption{The NUTS criterion successfully identifies the first TPOLO even for more sophisticated choices of the mass matrix, such as the quadratic potential \eqref{gaussV} with $\mathbf{M} = \mathbf{\Sigma}^{-1}$ (Figure \ref{fig:gaussSHO}b).
\label{fig:gaussNUTS2}}
\vspace{5mm}
\end{figure*}

In the more general case the phase offsets lead to beating, and the criterion vanishes at the average of the turning points,
\begin{equation*}
t \approx \frac{T}{2} \left[ n + \left(1 - \frac{2 }{\pi} \frac{\sum_{j} \phi_{j} }{n} \right) \right], \, n \in \mathbb{Z}.
\end{equation*}
For simple harmonic motion the NUTS criterion identifies the first TPOLO along the trajectory, and hence the optimal integration time, for any EMHMC system (Figure \ref{fig:gaussNUTS1}, \ref{fig:gaussNUTS2}).

\begin{figure*}[]
\centering
\includegraphics[width=7in]{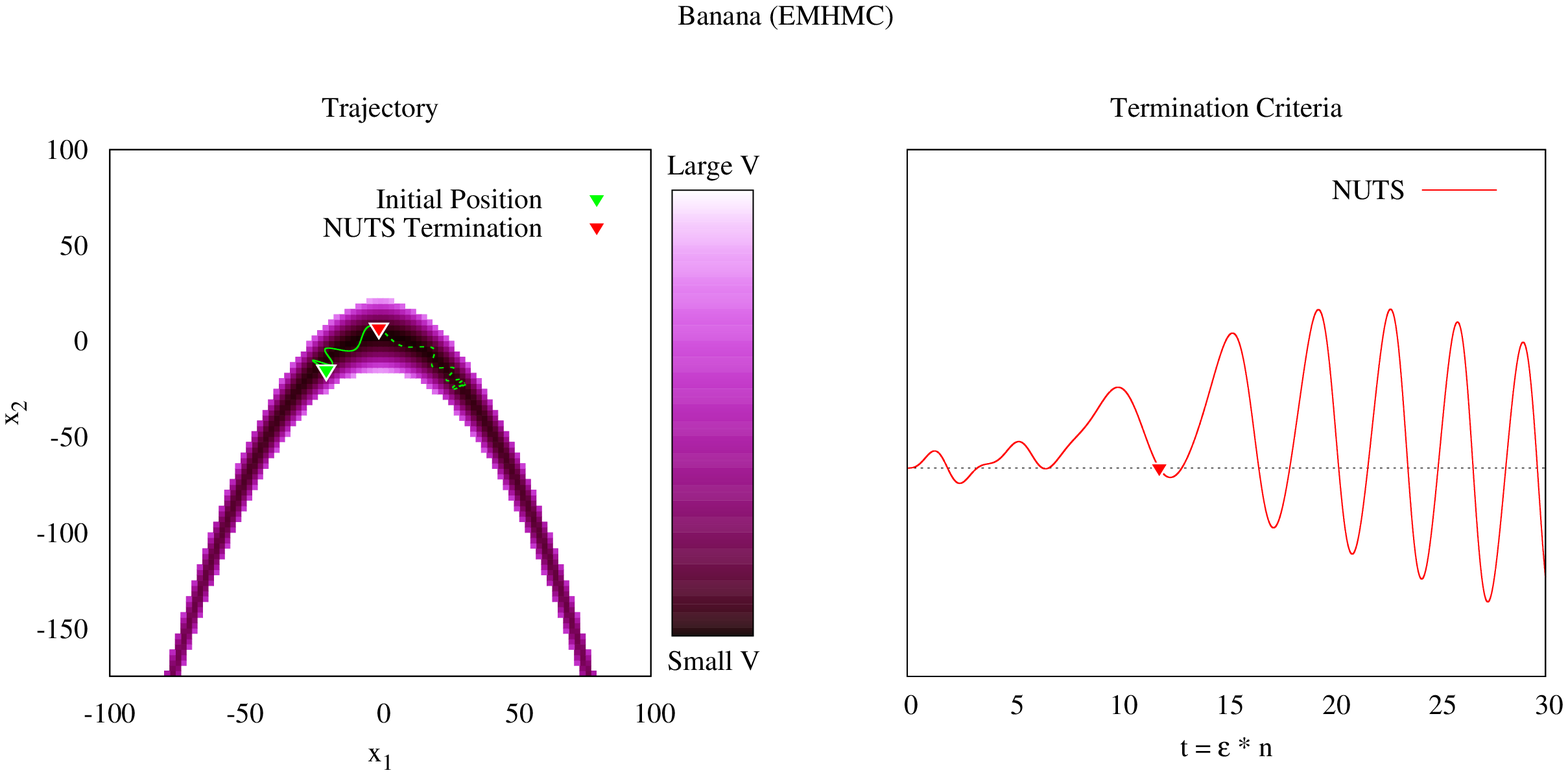}
\caption{On more complicated potentials, such as that arising from the banana distribution \eqref{banana}, the NUTS criterion terminates prematurely even when ignoring the transient behavior.
\label{fig:flatBananaNUTS}}
\vspace{5mm}
\end{figure*}

Although the NUTS criterion performs well in potentials that induce more complicated harmonic motion, it eventually begins to fail when applied to wells with more intricate geometry.  The potential from the \textit{banana} or \textit{twisted Gaussian} distribution~\cite{Haario1999},
\begin{equation} \label{banana}
V \! \left( R \right) = \frac{1}{2} \left[ \frac{ q_{1}^{2} \! \left( R \right) }{ \sigma_{1}^{2} } + \frac{ \left( q_{2}  \! \left( R \right) + \beta q_{1}^{2}  \! \left( R \right) - 100 \beta \right)^{2} }{ \sigma_{2}^{2} } \right]
\end{equation}
with $\beta = 0.03$, $\sigma_{1} = 0.01$, and $\sigma_{2} = 1$, for example, yields a non-convex well on which NUTS prematurely terminates (Figure \ref{fig:flatBananaNUTS}).

\begin{figure*}[]
\centering
\includegraphics[width=7in]{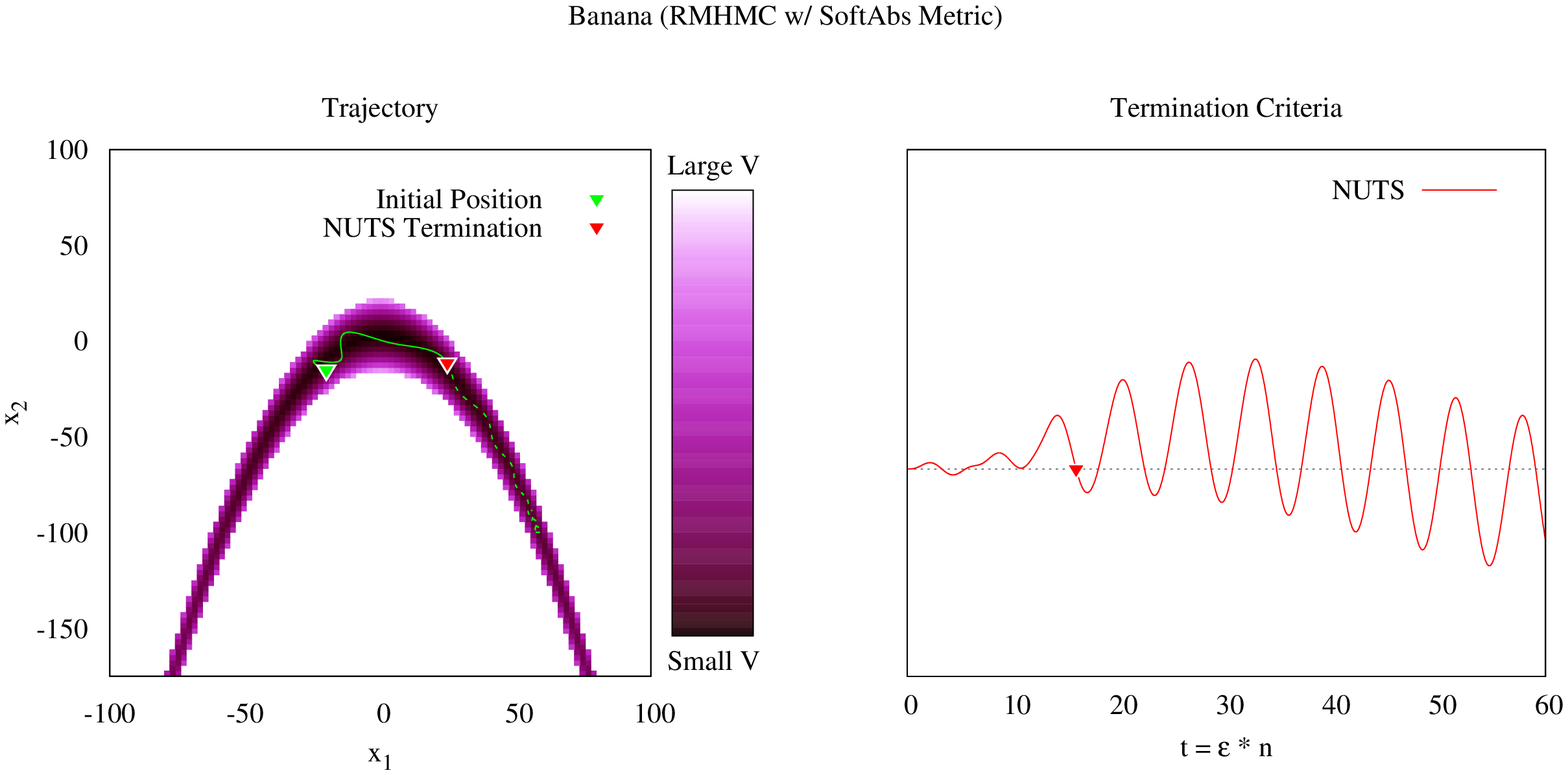}
\caption{RMHMC trajectories, here using the SoftAbs metric~\cite{Betan2012} with $\alpha = 1$, yield much smoother trajectories in the banana potential \eqref{banana}, but the ill-defined NUTS criterion still terminates prematurely.
\label{fig:softAbsBananaNUTS}}
\vspace{5mm}
\end{figure*}

\textit{Riemannian Manifiold Hamiltonian Monte Carlo} (RMHMC)~\cite{Girolami2011}
\begin{equation*}
H \! \left( S \right) = \frac{1}{2} \sum_{jk} p_{i} \! \left( S \right) p_{j} \! \left( S \right) \Lambda^{jk} \! \left( R \right) - \frac{1}{2} \log \! \left| \mathbf{\Lambda} \! \left( R \right) \right| + V \! \left( R \right),
\end{equation*}
endows the base manifold with a Riemannian geometry that can simplify the motion in such complex potentials, but here the distance $\mathbf{r}$ has no real meaning.  Although the trajectories are much smoother in RMHMC than EMHMC, the now ill-defined NUTS criterion still terminates prematurely (Figure \ref{fig:softAbsBananaNUTS}).

\section*{Identifying Turning Points on Riemannian Manifolds}

Although the motivating construction of the NUTS criterion required a measure of distance on the manifold, the intermediate form derived above does not.  In particular,
\begin{align*}
\left< \mathbf{p}  \! \left(R_t\right), \mbox{\boldmath{$\rho$}}  \! \left(R_t\right) \right>_{\mathbf{\Lambda} \left( R_t \right)}  &< 0
\end{align*}
with
\begin{align*}
\mbox{\boldmath{$\rho$}}  \! \left(R_t\right) = \frac{1}{t} \int_{0}^{t} \mathrm{d} \tau \, \mathbf{p} \! \left(R_\tau \right),
\end{align*}
is entirely well-defined on a Riemannian manifold, provided that we can add momenta from different points along the trajectory.  Fortunately, the symplectic geometry of $T^{*} \mathcal{M}$ furnishes such a means.

The \textit{canonical one-form}, $\theta \in T^{*} \left( T^{*} \mathcal{M} \right)$~\cite{Jose1998}, given in coordinates as
\begin{align*}
\theta &= \sum_{j} p_{j} \mathrm{d} q^{j},
\end{align*}
is a natural object on the cotangent bundle.  Note that $\theta$ has no $\mathrm{d} p_{j}$ components in any coordinate system --- it is a $\textit{horiztonal}$ form on $T^{*} \mathcal{M}$.

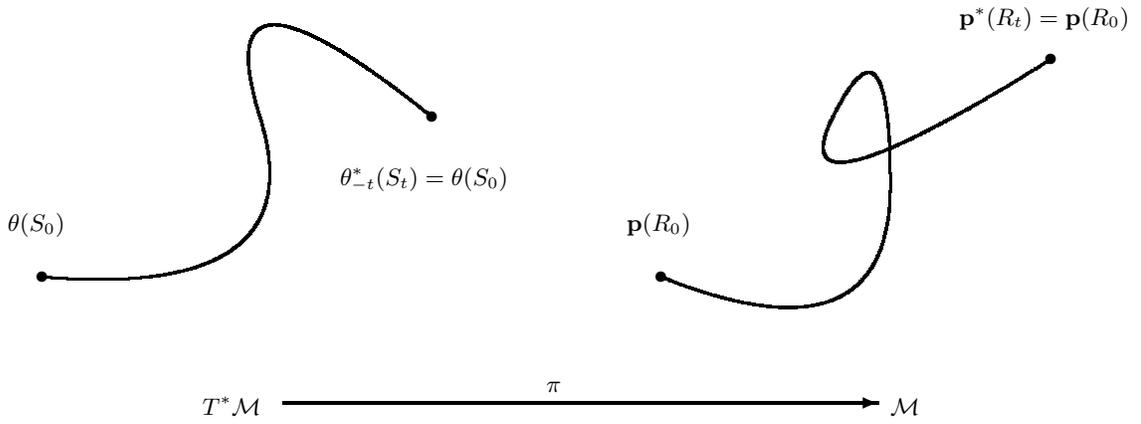
\begin{figure*}[t] 
\setlength{\unitlength}{0.12in} 
\centering
\begin{picture}(50, 20)
%
\put(10, 2){$T^{*} \mathcal{M}$}
\thicklines
\qbezier(3, 8)(15,7)(12.5,15)
\qbezier(12.5,15)(10,23)(20,15)
\put(3,8){\circle*{0.5}}
\put(20,15){\circle*{0.5}}
\put(1.5, 10) {$\theta \! \left( S_{0} \right)$}
\put(16, 12) {$\theta^{*}_{-t} \! \left( S_{t} \right) = \theta \! \left( S_{0} \right)$}
%
\put(40, 2){$\mathcal{M}$}
\thicklines
\qbezier(30, 8)(40,4)(40, 12)
\qbezier(40,12)(40,20)(37.5,15)
\qbezier(37.5,15)(35,10)(47,17.5)
\put(30,8){\circle*{0.5}}
\put(47,17.5){\circle*{0.5}}
\put(28.5, 10) {$\mathbf{p} \! \left( R_{0} \right)$}
\put(43, 19) {$\mathbf{p}^{*} \! \left( R_{t} \right) = \mathbf{p} \! \left( R_{0} \right)$}
\put(13.5, 2.5){\vector(1,0){26}}
\put(25, 3){$\pi$}
\end{picture} 
\caption{The Hamiltonian flow allows us to Lie drag the canonical one form, $\theta$, from any point $S_{0}$ to some final point $S_{t}$ along the trajectory.  This defines a unique momentum on the base manifold, $\mathcal{M}$, providing a map of the momentum at $R_{0} = \pi \! \left( S_{0} \right)$ to $R_{t} = \pi \! \left( S_{t} \right)$, where it can be compared with the current momentum.}
\label{fig:pDragging}
\end{figure*}

Now the canonical one-form can be \textit{Lie dragged} along the Hamiltonian flow~\cite{Schutz1980},
\begin{align*}
\theta^{*}_{t} &= \theta + \int_{0}^{t} \mathrm{d} \tau \, \mathcal{L}_{\vec{H}} \theta,
\end{align*}
where the components of the Lie derivative, $\mathcal{L}_{\vec{H}} \theta$, are given by
\begin{align*}
\left( \mathcal{L}_{\vec{H}} \theta \right)_{j} &= \sum_{k} \left[ \frac{ d q^{k} }{dt }  \frac{ \partial }{ \partial q^{k} } + \frac{ d p_{k} }{dt }  \frac{ \partial }{ \partial p_{k} } \right] \theta_{j} + \theta_{k}  \frac{ \partial }{ \partial q^{j} } \frac{ d q^{k} }{dt } \\
&=  \sum_{k} \left[ \frac{ d q^{k} }{dt }  \frac{ \partial }{ \partial q^{k} } + \frac{ d p_{k} }{dt }  \frac{ \partial }{ \partial p_{k} } \right] p_{j} + p_{k}  \frac{ \partial }{ \partial q^{j} } \frac{ d q^{k} }{dt } \\
&=  \sum_{k} \frac{ d q^{k} }{dt }  \frac{ \partial p_{j} }{ \partial q^{k} } + \frac{ d p_{k} }{dt }  \frac{ \partial p_{j} }{ \partial p_{k} } + p_{k}  \frac{d}{dt} \frac{ \partial q^{k} }{ \partial q^{j} } \\
&=  \sum_{k} \frac{ d q^{k} }{dt } 0 + \frac{ d p_{k} }{dt } \delta^{k}_{j} + p_{k}  \frac{d}{dt} \delta^{k}_{j} \\
&=  \sum_{k} \frac{ d q^{k} }{dt } 0 + \frac{ d p_{k} }{dt } \delta^{k}_{j} + p_{k}  0 \\
&= \frac{ d p_{j} }{dt }.
\end{align*}
Dragging $\theta$ from the beginning of a trajectory to the end then yields\footnote{Lie dragging is canonically defined as against the flow of the vector field, from the end of a trajectory to the beginning.  Here we're dragging with the flow, hence the the minus sign.}
\begin{align*}
\theta^{*}_{-t} \! \left( S_{t} \right) &= \sum_{j} \left[ p_{j} \! \left( S_{t} \right) + \int_{t}^{0} \mathrm{d}t \, \frac{ d p_{j} }{dt } \right] \mathrm{d} q^{j} \! \left( S_{0} \right) \\
&= \left[ p_{j} \! \left( S_{t} \right)  + p_{j} \! \left( S_{0} \right)  - p_{j} \! \left( S_{t} \right)  \right] \mathrm{d} q^{j}\! \left( S_{t} \right) \\
&= p_{j} \! \left( S_{0} \right)  \mathrm{d} q^{j}\! \left( S_{t} \right).
\end{align*}
Because $\theta$ is a horizontal form, $\theta^{*}_{-t}$ defines a unique covector $\mathbf{p}^{*} \! \left( R_{t} \right)$, with components 
\begin{align*}
p^{*}_{j} \! \left( R_{t} \right) &= \left( \theta^{*}_{t} \! \left( S_{t} \right) \right)_{j} \\
&= p_{j} \! \left( R_{0} \right),
\end{align*}
on the base manifold at $R_{t} = \pi \! \left( S_{t} \right) \in \mathcal{M}$.  This then defines a mapping of the momenta at any point $R_{0}$ along a trajectory to the $R_{0}$ (Figure \ref{fig:pDragging}).

We can now define $\mbox{\boldmath{$\rho$}}  \! \left(R_t\right)$ for any RMHMC system as
\begin{align*}
\mbox{\boldmath{$\rho$}}  \! \left(R_t\right) = \frac{1}{t} \int_{0}^{t} \mathrm{d} \tau \, \mathbf{p}^{*} \! \left(R_\tau \right),
\end{align*}
yielding the generalized NUTS criterion,
\begin{align} \label{gNUTS}
\left< \mathbf{p} \! \left(R_t\right), \mbox{\boldmath{$\rho$}}  \! \left(R_t\right) \right>_{\mathbf{\Lambda} \left( R_t \right)}  &< 0,
\end{align}
amenable to HMC on Riemannian manifolds.

In practice, the integral can be approximated by aggregating the momenta along discrete time intervals, $t_{k} - t_{k - 1} =  \delta t_{k}$,
\begin{equation*}
\mbox{\boldmath{$\rho$}} = \frac{1}{ \sum_{k} \delta t_{k} } \sum_{k} \mathbf{p}^{*} \! \left( R_{t_{k}} \right),
\end{equation*}
such as those naturally introduced in any discrete integrator.

\begin{figure*}[]
\centering
\includegraphics[width=7in]{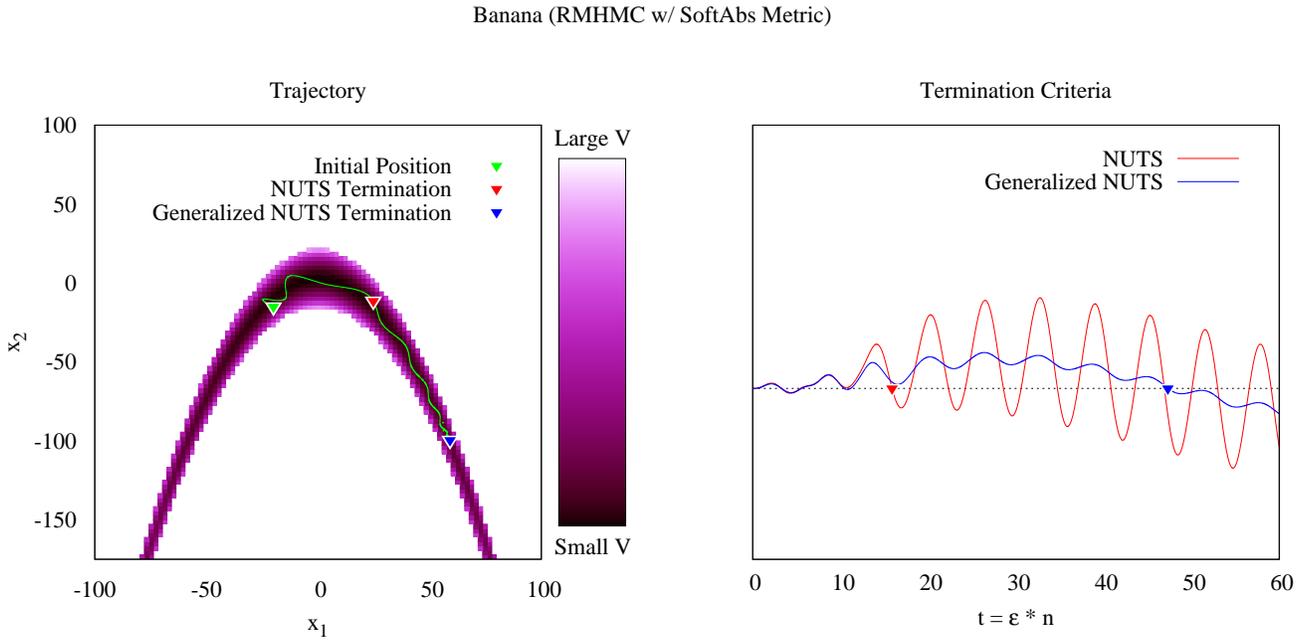}
\caption{Only with a well-defined criterion \eqref{gNUTS} do RMHMC trajectories, here using the SoftAbs metric~\cite{Betan2012} with $\alpha = 1$, terminate at the TPOLO of the banana distribution.  Note that the transient behavior persists, and care must still be taken at the beginning of a trajectory.
\label{fig:softAbsBananaGeoNUTS}}
\vspace{5mm}
\end{figure*}

Rigorously defined, this generalized criterion identifies the TPOLO even in warped distributions such as the banana (Figure \ref{fig:softAbsBananaGeoNUTS}).  

\vspace{10mm}

\section*{Conclusion}

By taking advantage of both symplectic and Riemannian geometry, NUTS is readily generalized to RMHMC.  When applied to bound trajectories, this criterion identifies the turning points of the longest oscillation and terminates the motion before it begins to double back on itself and waste computation.  

This generalization is readily adapted into the NUTS framework and is currently being in implemented in Stan~\cite{Stan2012}.

\section*{Acknowledgements}

I thank Andrew Gelman and the Stan development team for many engaging conversations, and Tarun Chitra for helpful discussion.

\vspace{50mm}

\bibliography{gnuts}{}

\end{document}